\documentclass[12pt]{article}
\usepackage{amssymb}
\usepackage{epsfig}
\begin{document}

\begin{titlepage}
\begin{flushright}
{\tt hep-ph/0307397}\\
OSU-HEP-03-11\\
MCTP-03-38\\
FERMILAB-Pub-03/222\\
August 2003\\
\end{flushright}
\vskip 2cm
\begin{center}
{\Large\bf \textbf Test of Gauge-Yukawa Unification} \vskip 1cm
{\normalsize\bf
Ilia Gogoladze,$^{a}$\footnote{On leave of absence from:
Andronikashvili Institute of Physics, GAS, 380077 Tbilisi,
Georgia.   E-mail address: ilia@hep.phy.okstate.edu}
Yukihiro Mimura,$^{b}$\footnote{E-mail address: mimura2y@uregina.ca}
S. Nandi$^{a,c}$\footnote{Summer visitor at Fermilab. E-mail address:
shaown@okstate.edu} and 
Kazuhiro Tobe$^{d}$\footnote{E-mail address: ktobe@umich.edu}}

\vskip 0.5cm
{\it $^a$Department of Physics, Oklahoma State University,
Stillwater,
OK~74078-3072 \\
$^b$Department of Physics, University of Regina, Regina, SK,
S4S 0A2, Canada\\
$^c$Fermi National Accelerator Laboratory, PO Box 500, Batavia, IL 60510\\
$^d$MCTP, Department of Physics, University of Michigan, Ann Arbor, MI
48109 \\  $^d$Department of Physics, University of California, Davis, CA
95616  \vskip 2cm}

\end{center}

\begin{abstract}
Recently it has been proposed that, in the framework of quantum
field theory, both the Standard Model gauge and Yukawa
interactions arise from a single gauge interaction in higher
dimensions with supersymmetry. This leads to the unification of
the Standard Model gauge couplings and the third family Yukawa couplings at
the GUT scale. In this work, we make a detailed study of this
unification using the current experimental data, and find a good
agreement in a significant region of the parameter space.
Similar 
relations, required in Finite Grand Unification models, are also studied.

\end{abstract}

\end{titlepage}

\section{Introduction}

Standard Model (SM), based on the gauge symmetry group,
$SU(3)_C\times SU(2)_L\times U(1)_Y$, has been very successful
experimentally. There is still no evidence beyond SM, except
possibly the neutrino masses and mixings. However, SM has many
parameters, such as three gauge  couplings, $g_3$, $g_2$ and $g_1$
and many Yukawa couplings such as $y_t$, $y_b$, $y_{\tau}$, $y_c$,
$y_s$, $y_\mu$, $y_u$, $y_d$, $y_e$ (where $y_t$ denotes the
Yukawa coupling of the top quark to the SM Higgs boson and so on),
mixing angles and phases.  It will be nice to relate some of these
parameters using symmetry.  Grand Unification Theory (GUT) such as
$SU(5)$, $SO(10)$ or $E_6$ relates the gauge couplings, since all the
gauge interactions of the SM arise from the single  gauge
interaction  of the unifying group.  This gives $g_3=g_2=g_1$ at
the unification scale, $M_{GUT}$, leading to the successful
prediction for the $\sin^2\theta_W$ at low energy in supersymmetric
(SUSY) GUT. For  specific choices of the Higgs sector, GUT can also
relates some of the Yukawa couplings. For example, in $SU(5)$ theory, we can
have $y_b=y_{\tau}$, where in $SO(10)$, we can have
$y_t=y_b=y_{\tau}$ at the GUT scale.  Such GUT relations among the
Yukawa coupling also lead to successful prediction at the low
energy for a significant region of the SUSY parameter space. (For
recent progress for top-bottom-tau Yukawa unification, see
Refs.~\cite{Baer:1999mc,Blazek:2001sb,Tobe:2003bc,Auto:2003ys,
Balazs:2003mm,Dermisek:2003vn}.)

One interesting fact we have observed is that in a scenario of
top-bottom-tau Yukawa unification together with gauge coupling
unification at the GUT scale, the unified Yukawa coupling ($y_G$)
can be very close to the unified gauge coupling ($g_G$)
($g_G\simeq 0.7$ and $y_G\geq 0.5$) in a wide range of parameter
space. This fact interestingly implies that the  origin of
Yukawa couplings might be related to the unified gauge
coupling. Therefore it naturally leads us to consider an interesting
possibility of ``gauge-Yukawa unification'' at high-energy scale. 
In this work, we will study the numerical test of such possibility.
We consider two different models in which gauge and Yukawa couplings
are related.
One is the higher dimensional model, and the other is so-called
finite GUT model.

In higher dimensional models, the Yukawa interactions can be
just part of the gauge interactions.
If we go to higher dimensions, the higher dimensional components
of the gauge bosons (say $A_5, A_6, \cdots$) are scalar fields,
and can be identified 
with the Higgs bosons~\cite{Manton:1979kb,Burdman:2002se,Hall:2001zb,Gogoladze:2003ci}.
The higher dimensional fermions include both chiral two-component spinors
in the four dimensional (4D) language. 
By orbifolding condition, the resulting 4D theories
can be chiral~\cite{Candelas:en}. The higher dimensional kinetic term of the fermion
includes the Dirac-type mass term of the Kaluza-Klein excited
modes such as $\overline{\Psi_L} \partial_5 \Psi_R$. The extra
dimensional derivative $\partial_5$ must be gauge covariant due to
the gauge invariance, and thus the lagrangian has the Yukawa term
such as $\overline{\Psi_L} A_5 \Psi_R$. Therefore, if the Higgs
fields which break electroweak symmetry are unified to the gauge
bosons in higher dimensions and the quarks and leptons are
zero-modes of the higher dimensional fermions, the Yukawa
interaction in the SM is just part of the gauge interactions. In
non-SUSY models, we need at least 6D to unify the standard model
Higgs fields with the higher dimensional components of gauge
bosons.
 The reason we cannot realize the unification in 5D is that
we need, at least, two real components to identify the higher dimensional
components of the gauge fields  with Higgs fields.
 In SUSY models,
we can construct gauge-Higgs unified models in 5D \cite{Burdman:2002se}. 
The 5D N=1 SUSY model corresponds to 4D N=2
SUSY. The N=2 gauge multiplet includes N=1 chiral superfield
$\Sigma$ (the imaginary part of its scalar components is $A_5$),
and we can identify the part of $\Sigma$ with the Higgs field. In
this model, thus, the Yukawa couplings can originate from the
gauge interaction. We can also construct gauge-Higgs unified models in 6D N=2
SUSY \cite{Hall:2001zb,Gogoladze:2003ci} which corresponds to 4D N=4 SUSY. The
N=4 gauge multiplet contains N=1 vector multiplet and three chiral
superfields.
 In the models of  Ref. \cite{Gogoladze:2003ci}, gauge fields, Higgs bosons
as well as the third family  matter  fermions are unified in a single multiplet belonging
to the adjoint representation of the unified gauge group in 6D.
In this way  both the Yukawa and the gauge interactions, in the compactified
4D theory, arise from a  single gauge interaction in 6D,
and thus the gauge and third family Yukawa couplings are unified at the 
compactification scale.
 The smallness of the first- and second-family Yukawa
couplings can be realized by using the volume suppression, fermion
localization \cite{Arkani-Hamed:1999dc}, 
Froggatt-Nielsen like mechanism \cite{Froggatt:1978nt}, and so on. 

The object of this work is to perform a detailed analysis of such
unification of the gauge couplings ($g_1$, $g_2$, $g_3$) and the
third family Yukawa couplings ($y_t$, $y_b$, $y_{\tau}$)
within the framework of SUSY models. We find
that a significant region of the parameter space allow such an
unification with the key prediction for $\tan\beta$ ($\tan\beta \simeq$ 52) and
the correlation among SUSY threshold corrections at low
energy.~\footnote{
In Ref.~\cite{Chkareuli:1998wi}, the naive calculation of gauge-Yukawa
unification has been performed.} Therefore, precise measurement of SUSY
parameters in future experiments will be quite important to test this
prediction of the gauge-Yukawa unification.

The relations between the Yukawa couplings and gauge couplings have
 also been found at the finite GUT models \cite{fin,beg},
which are four-dimensional models.
Though such relations do not arise from a symmetry, imposition of such relations at the
GUT scale lead to the finite GUT models, thus reducing the number
of parameters in the theory. In this work, we also investigate how
well such finite GUT relations work, and again find a good
agreement for a significant region of the parameter space.

\section{Formalism}

\subsection{Gauge-Yukawa unification}
A model realizing the unification of the gauge couplings ($g_1$,
$g_2$, $g_3$) and the third family Yukawa couplings ($y_t$, $y_b$,
$y_{\tau}$) was presented in \cite{Gogoladze:2003ci}. It has an $SU(8)$
gauge
symmetry in 6D with N=2 SUSY. N=2 SUSY in 6D
corresponds to N=4 SUSY in 4D, thus only the gauge
multiplet can be introduced in the bulk. 6D N=2 gauge multiplet,
expressed in terms of 4D, N=4 gauge multiplet, contains the vector multiplet
$V(A_{\mu}, \lambda)$ and three chiral multiplets in the adjoint
(63-dimensional) representation of the gauge group.  The
63-dimensional gauge  multiplet contains the gauge bosons (and
their superpartners) while the three 63-dimensional chiral
multiplets contain the third family matter fermions and the Higgs
bosons plus their superpartners. Two extra dimensions are
compactified in $T^2/Z_6$ orbifold. With suitable choice of the
$Z_6$ transformation matrix, $SU(8)$ is broken to the $SU(4)
\times SU(2)_L\times SU(2)_R\times U(1)^2$, and the theory reduces
to 4D N=1 SUSY Pati-Salam model with two extra $U(1)$
symmetry. The massless modes after compactification are the
Pati-Salam gauge fields, $\mathbf{(15, 1, 1), (1, 3, 1), (1, 1, 3)}$ plus two
additional singlet vector fields  $\mathbf{(1, 1, 1)}$ and $\mathbf{(1, 1, 1)}$, third-family
matter fermions  $\Psi_L=\mathbf{(4,\, 2,\, 1)}_{2,\,0}$ and $\Psi_{\bar R}=\mathbf{(\bar{4},\,
1,\, 2)}_{-2,\, -4}$ and bi-doublet Higgs fields, $H_1=\mathbf{(1,\, 2, \,
2)}_{0,\, 4}$ and $H_2=\mathbf{(1,\, 2, \, 2)}_{0,\,-4}$. Since all the
fields are contained in one representation of one simple gauge
symmetry (63-dimensional representation  of $SU(8)$ in 6D in this
case), all interactions in this theory arises only from gauge
interaction. The trilinear coupling for the chiral multiplets
\begin{equation}\label{aa1}
S=\int d^6 x \left[\int d^2 \theta \,
2\, 
{\rm
Tr}\left(-\sqrt{2} g_6 \Sigma [\Phi, \Phi ^c]\right)+h.c.\right]
\end{equation}
includes the Yukawa interaction terms
\begin{equation}\label{aa2}
S=\int d^6 x \int d^2 \theta \,y_6 {\Psi}_L H_1 \Psi_{\bar R}
+h.c.
\end{equation}
In Eq. (\ref{aa1}), $\Sigma, \, \Phi,\, \Phi^c$  are chiral
multiplets containing the third family chiral fields, $\Psi_L$ and $\Psi_{\bar R}$, and the
bi-doublet Higgs fields, $H_1$ and $H_2$, and $g_6$ and $y_6$ are the 6D gauge
and Yukawa couplings. Eqs. (\ref{aa1}) and (\ref{aa2}) leads to
$g_6=y_6$ with proper kinetic normalization. Integrating out the two extra dimensions, we obtain
$y_4=g_4$ for the 4D coupling leading to
\begin{equation}\label{aa3}
g_1=g_2=g_3=y_t=y_b=y_{\tau}(=y_{\nu_\tau}^{\rm Dirac})
\end{equation}
at the compactification scale ($M_c$) which is also the unification scale
in our theory.
We assume that the Pati-Salam symmetry, as well as the two extra $U(1)$ are
broken at $M_c$ to the $SU(3)_c\times
SU(2)_L \times U(1)_Y$  using suitable Higgs fields at the brane
so that the particle spectrum below $M_c$ is the same as
in MSSM. 
This model is one concrete example which predicts the relation (\ref{aa3}).
The 6D N=2 SUSY $SU(8)$ model can be modified to 6D N=2 SUSY 
$SO(16)$~\cite{Gogoladze:2003ci}, and many other models can be constructed
with different low energy symmetry, but all having gauge-Yukawa unification
\cite{big_paper}.
The relation (\ref{aa3}) is the gauge and Yukawa
unification for the third family, whose validity and phenomenological
implication  will be tested in the next section.

\subsection{Finite GUT unification}

Another possibility to connect the Gauge and the Yukawa couplings  is finite
N=1 SUSY theory
  \cite{fin,beg}
 wherein the $\beta$-functions for the
gauge and the Yukawa couplings vanish to all orders in
perturbation theory.
 In order to have all loop finite theory,
there is definite set of conditions which need to  be satisfied.
Below we  briefly review these  conditions .

The one-loop gauge and Yukawa $\beta$-functions and the one-loop
anomalous dimension of the matter fields in a generic SUSY
Yang-Mills theory are given by \cite{PW}:
\begin{eqnarray}\label{bg1}\beta^{(1)}_{g}&=&\frac{g^3}{16\pi^{2}}\left
(\sum_{R}
    T(R)-3C_{2}(G)\right),\\
\beta^{(1)}_{ijk}&=&\lambda_{ijp}\gamma^{(1)}{}^{p}_{k}
+(k\leftrightarrow i)+(k\leftrightarrow j),
\\
\gamma^{(1)}{}^i_{j}&=&\frac1{16\pi^2}[\lambda^{ikl}\lambda_{jkl}-2C_{2}(R)g^{2}\delta^{i}_{j}],
\end{eqnarray}
where $T(R)$, $C_{2}(R)$ and $C_{2}(G)$ are the Dynkin
indices for  the matter fields and the quadratic Casimirs for  the
matter and gauge representations respectively. $\lambda^{ijk}$ and
$\beta^{(1)}_{ijk}$ are the Yukawa couplings and the one-loop
Yukawa $\beta$-functions  of $\lambda^{ijk}$. The criteria of all
loop finiteness for N=1 SUSY gauge theories can be
stated as follows \cite{LPS}:

(I) It should be free from gauge anomaly.

(I$\!$I) The gauge $\beta$-function vanishes at one loop:
$\beta^{(1)}_{g}=0$.

(I$\!$I$\!$I) There exists solution of the form $\lambda=\lambda(g)$ for
the conditions of vanishing one-loop anomalous dimensions:
$\gamma^{(1)}{}^i_{j}=0$.

(I$\!$V) The solution is isolated and non-degenerate when
considered as a solution of vanishing one-loop Yukawa
$\beta$-function:
$\beta^{(1)}_{ijk}=0$.
\newline
If all four conditions are satisfied, the dimensionless parameters
of the theory would depend on a single gauge coupling constant and
the $\beta$-functions will vanish to all orders.

Models  satisfying the criteria (I) through (I$\!$V) have been
found in the $SU(5)$ SUSY GUT \cite{fin,beg} with appropriate
particle contents. One such solution \cite{beg}  relating the
gauge and the third family Yukawa couplings based $SU(5)\times
A_4$ symmetry is
\begin{equation}
\label{finite_gut_relation}
 y_b=y_\tau=\frac{\sqrt{3}}{2} y_t={\frac{3}{\sqrt{10}}}
g_G\,,
\end{equation}
 where $A_4$ is the group of even permutation~\cite{Ma}, $y_t$, $y_b$
and $y_\tau$ are the  top, bottom and tau Yukawa 
couplings, and $g_G$ is the gauge coupling at the unification
scale.

\section{Analysis of  gauge-Yukawa unification scenarios}

In this section, we analyze two gauge-Yukawa unification scenarios in
which Yukawa couplings can be related to the unified gauge coupling:
``gauge-Yukawa unification ($y_t=y_b=y_\tau=g_G$)'' and
``finite-GUT unification''.
It has been stressed that high-energy Yukawa couplings are highly
sensitive to low-energy SUSY threshold corrections to Yukawa and
gauge couplings especially in large $\tan\beta$ case.
Therefore, in a study of any Yukawa unification scenarios,
an inclusion of low-energy SUSY threshold corrections is very important.
Following the analysis done in Ref.~\cite{Tobe:2003bc},
we perform a semi-SUSY model-independent analysis to see if
the gauge-Yukawa unification scenarios are realistic or not.
In our analysis, we use a dimensional reduction ($\overline{\rm DR}$)
renormalization scheme, which is known to be consistent with SUSY.
${\overline{\rm{DR}}}$ Yukawa couplings ($y_{t,b,\tau}$)
and gauge couplings ($g_i$) in the MSSM at Z-boson mass scale
are written as follows:
\begin{eqnarray}
{y}_t(m_Z) &=&
\frac{\sqrt{2}\bar{m}_t^{\rm{MSSM}}(m_Z)}{\bar{v}(m_Z) \sin\beta}
=\frac{\sqrt{2} \bar{m}_t^{\rm{SM}}(m_Z)}{\bar{v}(m_Z) \sin\beta}
\left(1+\delta_t\right),\\
{y}_{b,\tau}(m_Z) &=&
\frac{\sqrt{2}\bar{m}_{b,\tau}^{\rm{MSSM}}(m_Z)}{\bar{v}(m_Z) \cos\beta}
=\frac{\sqrt{2}\bar{m}_{b,\tau}^{\rm{SM}}(m_Z)}{\bar{v}(m_Z) \cos\beta}
\left(1+\delta_{b,\tau}\right),\\
{g}_i(m_Z)&=&
\bar{g}_i^{\rm{SM}}(m_Z)\left(1+\delta_{g_i}\right),~~(i=1-3)
\end{eqnarray}
where $\bar{m}_i^{\rm{SM}}$ and $\bar{g}_i^{\rm{SM}}$ are
$\overline{\rm{DR}}$ quantities defined in the SM, and
$\bar{v}$ and $\tan\beta$ are $\overline{\rm DR}$ values in the MSSM.
They are determined following the analysis in Ref.~\cite{Tobe:2003bc}.
(See Ref.~\cite{Tobe:2003bc} for detail and references.)
Especially when we calculate $\bar{m}_i^{\rm{SM}}(m_Z)$,
we adopt top pole mass ($m_t=174.3\pm 5.1$ GeV), tau pole mass
($m_\tau=1776.99 ^{+0.29}_{-0.26}$ MeV) and $\overline{\rm{MS}}$
bottom mass ($\bar{m}^{\rm MS}_b(\bar{m}^{\rm MS}_b)
=4.26\pm0.30$ GeV).
The quantities $\delta_{t,b,\tau,g_i}$ represent SUSY threshold
corrections. If we choose a certain SUSY breaking scenario, they
are fixed. In our analysis, however, we treat them as free parameters
without specifying any particular SUSY breaking scenario.\footnote{
There are several known SUSY breaking mechanisms.
However, we do not know whether known mechanisms are really realized
in nature. Therefore, we believe that at this stage our SUSY
model-independent analysis is the most appropriate approach to
investigate gauge-Yukawa unification scenarios.}

When all parameters $\delta_{t,b,\tau,g_i}$ are specified, all
$\overline{\rm{DR}}$ couplings in the MSSM are determined at $m_Z$. Then
we use two-loop renormalization group equations (RGEs) for the MSSM
couplings in order to study the unification of couplings at the GUT scale.
Requiring a certain unification scenario, we can obtain
constraints among parameters $\delta_{t,b,\tau,g_i}$, as we will see later.
In this paper, we assume that the theory between $m_Z$ and the GUT scale
is well described by the MSSM.

\subsection{Gauge-Yukawa unification $y_t=y_b=y_\tau=g_G$}
Here we consider a possibility that all SM gauge
couplings, top, bottom and tau Yukawa couplings are unified
at the GUT scale, which we call ``gauge-Yukawa unification''
($y_t=y_b=y_\tau=g_G$).
In order to study the gauge-Yukawa unification,
first we look for a region where top, bottom and
tau Yukawa couplings are unified ($y_t=y_b=y_\tau\equiv y_G$) at the GUT
scale. We define the GUT scale ($M_G$) as a scale where
$g_1(M_G)=g_2(M_G)\equiv g_G$. In our analysis, we allow the possibility that  the strong
gauge coupling is not exactly unified:
$g_3(M_G)^2/g_G^2=1+\epsilon_3$
where $\epsilon_3$ can be a few \%. This mismatch $\epsilon_3$ from
exact unification can be considered to be due to a GUT scale threshold
correction to the unified gauge coupling.

\begin{figure}[p]
\centering
\includegraphics*[angle=0,width=13cm]{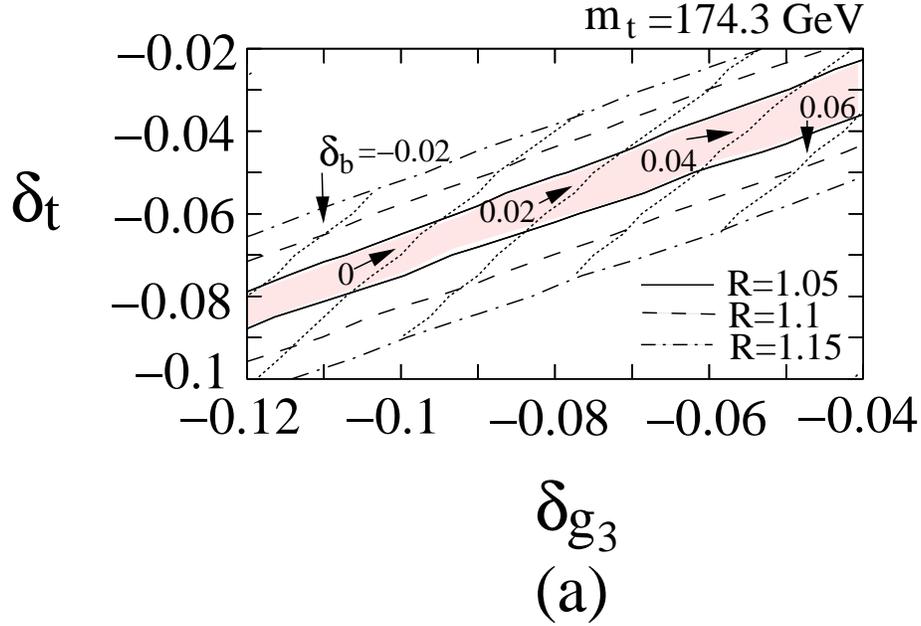}
\includegraphics*[angle=0,width=13cm]{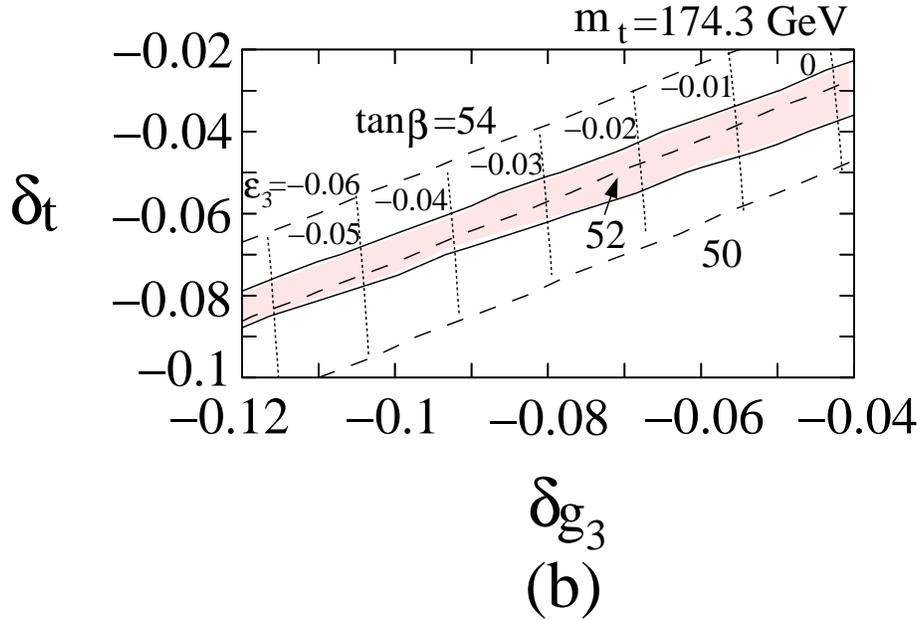}
\caption{Parameter space satisfying the gauge-Yukawa unification.
Contours of $\delta_b$ (dotted lines in Fig.~(a)), $\tan\beta$ (dashed lines
in Fig.~(b)) and $\epsilon_3$ (dotted lines in Fig.~(b)) are shown as a
function of $\delta_t$ and $\delta_{g_3}$, required for Yukawa unification
($y_t=y_b=y_\tau$). After finding the region for the Yukawa unification,
contours of a parameter $R$ (defined in text) are plotted in
Fig.~(a). The shaded regions represent a region where the gauge-Yukawa
unification is  achieved within $5\%$ level ($R \leq 1.05$).
Here we have fixed  $m_t=174.3$ GeV, $\bar{m}^{\rm MS}_b(\bar{m}^{\rm MS}_b)
=4.26$ GeV, $m_\tau=1776.99$ MeV, $\delta_\tau=0.02$,
$\delta_{g_1}=-0.006$ and $\delta_{g_2}=-0.02$.}
\label{gauge-yukawa_unif}
\end{figure}

In Fig.~\ref{gauge-yukawa_unif}, contours of $\delta_b$ (dotted
lines in Fig.~(a)), $\tan\beta$ (dashed lines in Fig.~(b)) and
$\epsilon_3$ (dotted lines in Fig.~(b))
are shown as a function of $\delta_t$ and $\delta_{g_3}$,
which are required for the Yukawa unification ($y_t=y_b=y_\tau$)
at the GUT scale.
In Fig.~\ref{gauge-yukawa_unif}, we take central values of input
fermion masses ($m_t=174.3$ GeV, $\bar{m}_b^{\rm MS}(\bar{m}_b^{\rm
MS})=4.26$ GeV
and $m_\tau=1776.99$ MeV) and $\delta_\tau=0.02$.
In order to fix $\delta_{g_{1,2}}$, we assume that all SUSY mass parameters
which contribute to $\delta_{g_{1,2}}$ are equal to $500$ GeV
($\delta_{g_1}=-0.006$ and $\delta_{g_2}=-0.02$).
As shown in Fig.~\ref{gauge-yukawa_unif}, 
$\tan\beta$ should be about $50$, and the value of
$\delta_b$ should be a few \%~\cite{Blazek:2001sb, Tobe:2003bc}, 
which is much smaller than one naively expected in large $\tan\beta$
case~\cite{Hempfling:1993kv}.

Our next question is: ``Is there any region where the unified
Yukawa coupling ($y_G$) is really unified into the unified gauge
coupling ($g_G$)?''
After requiring Yukawa unification, we calculate a parameter $R$
defined as follows:
\begin{eqnarray}
R&\equiv& \frac{{\rm max}(y_t,y_b,y_\tau,g_1,g_2,g_3)}
{{\rm min}(y_t,y_b,y_\tau,g_1,g_2,g_3)}
\simeq \left\{
\begin{array}{c}
y_G/g_G ~~{\rm for}~~y_G>g_G,\\
g_G/y_G ~~{\rm for}~~y_G<g_G.
\end{array}
\right.
\end{eqnarray}
When $R=1$, exact gauge-Yukawa unification happens.
In Fig.~\ref{gauge-yukawa_unif}(a), contours of $R$ are
shown to see if there is a region in which the gauge-Yukawa unification
happens.
As one can see from Fig.~\ref{gauge-yukawa_unif}(a),
there is a region where the gauge-Yukawa unification is well achieved.
In the shaded regions of Fig.~\ref{gauge-yukawa_unif}, the gauge-Yukawa
unification is realized within $5\%$ level ($R\leq1.05$) allowing
$\epsilon_3$ to be a few \%. Note that the gauge-Yukawa unification
requires an interesting relation between $\delta_t$ and $\delta_{g_3}$
and a very specific $\tan\beta$ ($\tan\beta\simeq 52$)
in addition to small $\delta_b$.
We have checked that the value of $\epsilon_3$ is quite sensitive to
values of $\delta_{g_{1,2}}$ because a change of $\delta_{g_{1,2}}$
shifts the unified gauge coupling $g_G$ but not
$g_3(M_G)$ very much. On the other hand, the relation between
$\delta_t$ and $\delta_{g_3}$ as well as the value of $\tan\beta$
does not depend on $\delta_{g_{1,2}}$ very much.
Therefore we have found that the relation between $\delta_t$
and $\delta_{g_3}$ and the values of $\tan\beta$ ($\tan\beta\simeq 52$)
are rather stable predictions from the gauge-Yukawa unification.
Thus in principle, if SUSY parameters were measured precisely enough
to know $\delta_t$, $\delta_{g_3}$ and $\tan\beta$,
the gauge-Yukawa unification could be tested.

In the above analysis,  we have fixed top and bottom masses.
We  note that a change of top (bottom) mass simply shifts an
allowed region of parameter $\delta_t$ ($\delta_b$).
For example, if we take $m_t$ to be $174.3+5.1$
($\simeq174.3(1+0.03)$) GeV, the allowed region of $\delta_t$ is shifted
by about $-0.03$. Since uncertainties of top and bottom masses are
still large, the precise determination of these masses is also quite
important to test the gauge-Yukawa unification.

We comment on some possible high-energy threshold corrections.
One possible correction would be due to neutrino Yukawa couplings.
If neutrino Yukawa couplings are large and run below the GUT scale,
they induce at most a few \% corrections to GUT-scale Yukawa couplings.
As a result, the effects modify the value of the  unified Yukawa coupling
and the relation among the SUSY threshold correction parameters by a few
\%, as discussed in Ref.~\cite{Tobe:2003bc}.
Other possible corrections could  originate  from the theory
of extra-dimensions~\cite{Kakushadze:1999bb}. There would be corrections
from the brane 
localized interactions. These corrections can be negligible if
the volume of extra dimensions is large. Also there might be some
corrections from the integration of extra-dimensions. These corrections,
however, highly depend on the nature of extra-dimensions (number of
extra-dimensions and SUSY, topology of extra-dimensions etc).
Therefore, we will not try to discuss the model-dependent corrections.
Instead, we can see some effects to the allowed region
for the gauge-Yukawa unification, adopting the parameter $R$.
We have plotted contours of $R$ in Fig.~\ref{gauge-yukawa_unif}(a).
These contours show how much the allowed region can change
if a deviation from $R=1$ originates from these high-energy threshold
corrections.
As can be seen, if the deviation is of the order of a few \%, still
the allowed region is well constrained.
In the  discussions
 in section 4,
 we will assume that the gauge-Yukawa unification
is realized within 5\% ($R\leq 1.05$) to see the implication of the
gauge-Yukawa unification.

\subsection{Finite GUT unification}
In this section, we consider another type of gauge-Yukawa unification.
In a model discussed in Ref.~\cite{fin,beg},
the finiteness condition implies the
unification at the GUT scale given by Eq.~(\ref{finite_gut_relation}).
This provides an interesting relation between Yukawa and gauge
couplings at the GUT scale, and we call it  ``finite GUT unification''.

\begin{figure}[p]
\centering
\includegraphics*[angle=0,width=12.9cm]{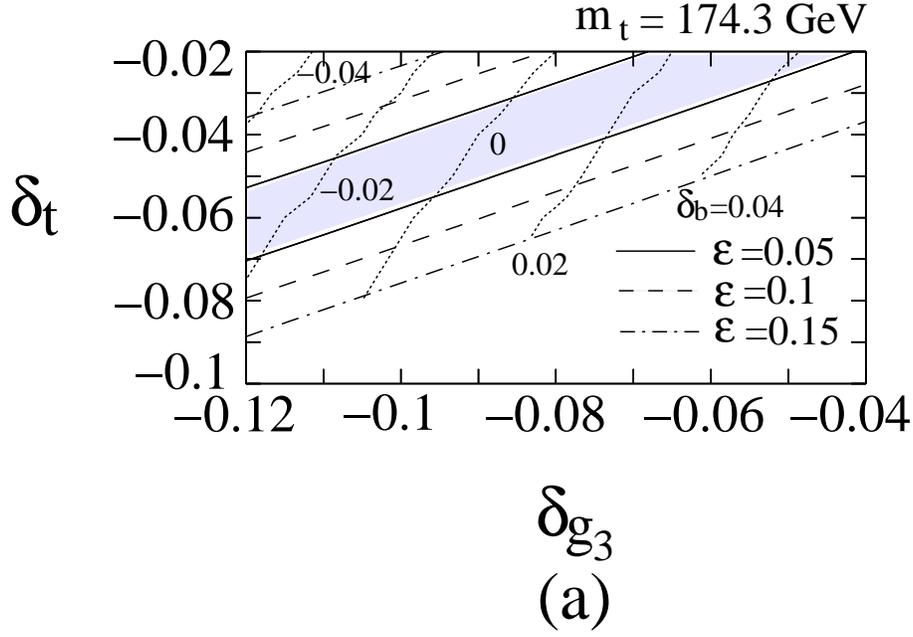}
\includegraphics*[angle=0,width=12.9cm]{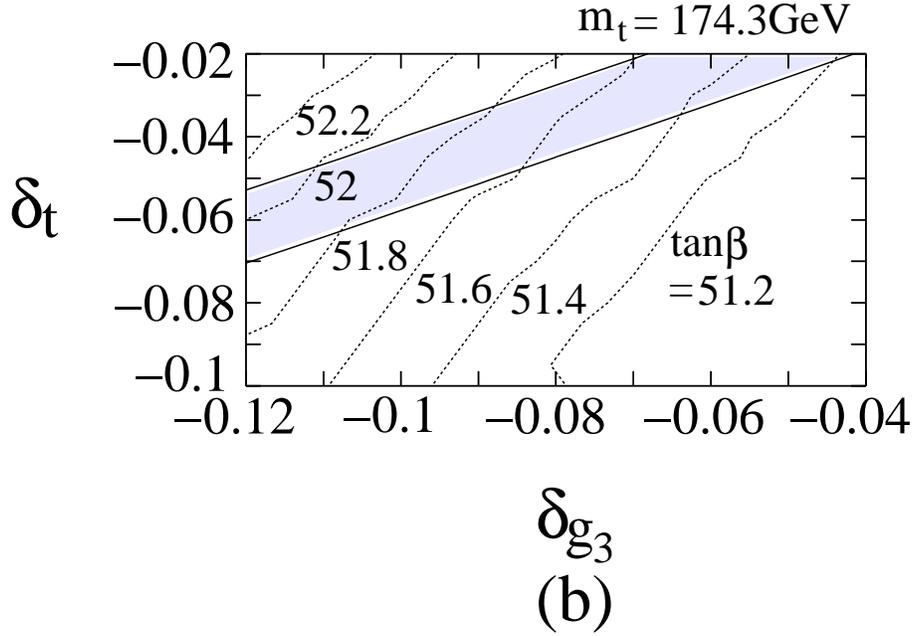}
\caption{Parameter space satisfying the finite GUT unification.
Contours of $\delta_b$ (dotted lines in Fig.~(a)) and $\tan\beta$
(dotted lines in Fig.~(b)) are shown as a function of $\delta_t$ and
$\delta_{g_3}$, required for bottom-tau-gauge unification
($y_b=y_\tau=g_G\sqrt{9/10}$). After finding the bottom-tau-gauge
unification, we also plot contours of
$\epsilon$ (defined in text) in Fig.~(a).
The shaded regions represent a region in which
the finite GUT gauge-Yukawa unification is achieved within
5\% level ($\epsilon \leq 0.05$). Here we have fixed
$m_t=174.3$ GeV, $\bar{m}^{\rm MS}_b(\bar{m}^{\rm MS}_b)
=4.26$ GeV, $m_\tau=1776.99$ MeV, $\delta_\tau=0.02$,
$\delta_{g_1}=-0.006$ and $\delta_{g_2}=-0.02$.}
\label{finite_GUT_unif}
\end{figure}

In order to find an allowed region for the finite GUT unification,
we first search for a region where bottom, tau and gauge
coupling unification in Eq.~(\ref{finite_gut_relation})
($y_b=y_\tau=\sqrt{9/10}g_G$) happens.
In Fig.~\ref{finite_GUT_unif}, we show relations among parameters
$\delta_t$, $\delta_{g_3}$, $\delta_b$ and $\tan\beta$
which are required for the bottom-tau-gauge unification in
Eq.~(\ref{finite_gut_relation}). Contours of $\delta_b$ (dotted
lines in Fig.~(a)) and $\tan\beta$ (dotted lines in Fig.~(b)) are
shown as a function of $\delta_t$ and $\delta_{g_3}$.
Here we have taken central values of input fermion masses, and
$\delta_\tau=0.02$, $\delta_{g_1}=-0.006$, $\delta_{g_2}=-0.02$.
Similar to the gauge-Yukawa unification discussed in the previous
section, $\delta_b$ is required to be small, and $\tan\beta$ should be
around 50.

Then we look for a region in which top and gauge coupling unification
in Eq.~(\ref{finite_gut_relation}) ($y_t=g_G\sqrt{6/5}$) is realized
after finding the bottom-tau-gauge unification in
Eq.(\ref{finite_gut_relation}).
We define a parameter $\epsilon$:
\begin{eqnarray}
\epsilon=\frac{\left| y_t-g_G \sqrt{6/5}\right|}{y_t},
\end{eqnarray}
so that $\epsilon=0$ if the finite GUT unification
Eq.~(\ref{finite_gut_relation}) is achieved.
In Fig.~\ref{finite_GUT_unif}(a),
we plot contours of $\epsilon$. As can be seen from
Fig.~\ref{finite_GUT_unif}, we found a region where the finite GUT
unification is realized. The shaded regions in
Fig.~\ref{finite_GUT_unif} represent a region where
the finite GUT gauge-Yukawa unification is achieved within
5\% level ($\epsilon \leq 0.05$).

Notice that the finite GUT gauge-Yukawa unification constrains SUSY
threshold  correction parameters $\delta_{t,b,\tau,g_i}$. Especially, it
requires a correlation between $\delta_t$ and
$\delta_{g_3}$, which interestingly suggests a slightly different
relation from  the one for the gauge-Yukawa unification discussed in the
previous section.

In the  next section, we discuss the implication of the relations between
$\delta_t$ and $\delta_{g_3}$ to SUSY mass spectrum.

\section{Implications to superparticle mass spectrum}

We have analyzed two different gauge-Yukawa unification scenarios.
Each scenario predicts a certain relation between $\delta_t$ and
$\delta_{g_3}$. It is interesting to discuss implications of these
relations to superparticle mass spectrum.
\begin{figure}[t]
\centering
\includegraphics*[angle=0,width=12cm]{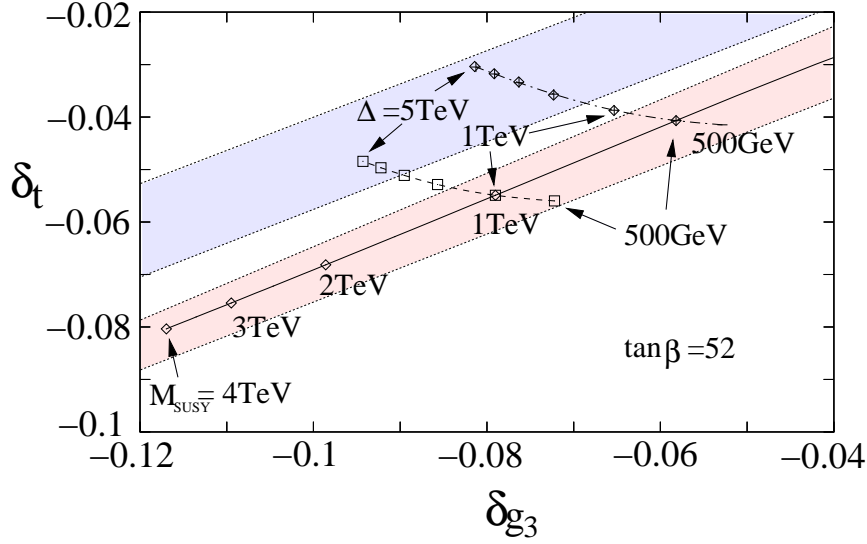}
\caption{Relations between $\delta_t$ and $\delta_{g_3}$.
In the solid line, all SUSY mass parameters are set to be equal to $M_{\rm
SUSY}$, then the relation between $\delta_t$ and $\delta_{g_3}$ is
shown as a function of $M_{\rm SUSY}$.
In the dashed and dash-dotted lines, all the first and second generation
squarks, wino and bino masses are assumed to be equal to $\Delta$ and
the rest of SUSY masses to be $M_{\rm SUSY}$.
In the dashed (dash-dotted) line, the relation between $\delta_t$ and
$\delta_{g_3}$ is shown for $M_{\rm SUSY}=1$ TeV
($M_{\rm SUSY}=500$ GeV) as a function of $\Delta$.
Two shaded regions represent the allowed regions for ``gauge-Yukawa
unification'' ($R\leq 1.05$) in lower shaded region and
for ``finite GUT unification'' ($\epsilon\leq0.05$) in upper shaded
region.
Here we have fixed $m_t=174.3$ GeV, $\bar{m}^{\rm MS}_b(\bar{m}^{\rm MS}_b)
=4.26$ GeV, $m_\tau=1776.99$ MeV and $\tan\beta=52$.}
\label{SUSY_relation}
\end{figure}

If all SUSY mass parameters (gaugino masses $M_{g_i}$,
sfermion masses $m_{\tilde{f}}$, and $\mu$-term) are simply set to be
equal to $M_{\rm SUSY}$,  and A-term is set  to be zero, and then we calculate
$\delta_t$ and $\delta_{g_3}$ as a function of $M_{\rm SUSY}$, we get a
relation between 
$\delta_t$ and $\delta_{g_3}$ as shown in Fig.~\ref{SUSY_relation}
(solid line). Here we have assumed $m_t=174.3$ GeV,
$\bar{m}_b^{\rm MS}(\bar{m}_b^{\rm MS})=4.26$ GeV, $m_\tau=1776.99$ MeV
and $\tan\beta=52$.
We also show points for $M_{\rm SUSY}=500$ GeV, 1, 2, 3 and
4 TeV on the solid line in Fig.~\ref{SUSY_relation}. One can see that
as $M_{\rm SUSY}$ gets larger, both $\delta_t$ and $\delta_{g_3}$ become
smaller. In Fig.~\ref{SUSY_relation}, two shaded regions
represent the allowed regions for ``gauge-Yukawa unification''
(lower shaded region) and for ``finite GUT unification'' (upper shaded
region) obtained in the previous section.
As can be seen from Fig.~\ref{SUSY_relation}, 
interestingly the solid line just lies
on the allowed region for the gauge-Yukawa unification. 
This choice of SUSY mass parameters is one example to realize the
relation between $\delta_t$ and $\delta_{g_3}$ suggested by the
gauge-Yukawa unification.
Therefore,
getting the relation predicted by the gauge-Yukawa unification
is not particularly difficult.

At given $\delta_t$,  the finite GUT unification
requires smaller $\delta_{g_3}$ than one for the gauge-Yukawa
unification. Note that all colored SUSY particles contribute to
$\delta_{g_3}$, on the other hand, only the third generation squarks
as well as gauginos and higgsinos contribute to $\delta_t$.
Thus it is suggested that the finite GUT unification
prefers the heavier first and second generation squarks more than the
gauge-Yukawa unification does. This is an interesting implication from
two different gauge-Yukawa unification scenarios.

We assume that all the first and second-generation squark masses,
wino and bino masses are equal to $\Delta$, and the rest of SUSY parameters
stays at $M_{\rm SUSY}$. Then we show how the relation between $\delta_t$ and
$\delta_{g_3}$ changes as a function of $\Delta$ in
Fig.~\ref{SUSY_relation}. Dashed line is for $M_{\rm SUSY}=1$ TeV, and
dash-dotted line for $M_{\rm SUSY}=500$ GeV. We also show points
for $\Delta=500$ GeV, 1, 2, 3, 4 and 5 TeV on both dashed and dash-dotted
lines in Fig.~\ref{SUSY_relation}.
One can see that clearly rather heavy first and second generation
squarks are preferred for the finite GUT unification.

In order to realize the gauge-Yukawa unification, one needs to satisfy
one more constraint on $\delta_b$. To get small $\delta_b$, a
cancellation or a suppression among contributions to $\delta_b$ is
needed as discussed in Ref.~\cite{Blazek:2001sb,Tobe:2003bc}.
Therefore, keeping the relation between $\delta_t$ and $\delta_{g_3}$,
we need to tune parameters such as stop, sbottom, gluino,
chargino masses and A-term to get the required $\delta_b$.
For example, in a case with $M_{\rm SUSY}=\Delta=500$ GeV (1 TeV) in
Fig.~\ref{SUSY_relation} for the gauge-Yukawa unification,
we need $M_{\tilde{g}_3}=500$ GeV (1 TeV),
$m_{\tilde{Q}_3}=m_{\tilde{t}_R}=200$ GeV ($400$ GeV),
$m_{\tilde{b}_R}=1500$ GeV (3 TeV), $\mu=100$ GeV and $A_t=0.45 M_{\rm 
SUSY}$ ($0.4 M_{\rm SUSY}$) to obtain small $\delta_b$ 
($\delta_b\sim 0.04$ ($0.03$)) keeping the
relation between $\delta_t$ and $\delta_{g_3}$.
Because of the relation between $\delta_t$ and $\delta_{g_3}$ and
the constraint on $\delta_b$ predicted by the gauge-Yukawa unification, 
SUSY mass parameters have to be highly correlated.
As realistic examples, we have noticed that in the supergravity-type SUSY
breaking scenario, data point 1 on Table 1 in the second paper 
of Ref.~\cite{Blazek:2001sb} and data points 1--3 in
Ref.~\cite{Dermisek:2003vn}
are explicit cases for the gauge-Yukawa unification.
Therefore, there exists a model which realizes the gauge-Yukawa
unification as well as provides the observed relic density of dark
matter and a good fit to precision electroweak data.~\footnote{
According to Ref.~\cite{Dermisek:2003vn}, it seems that all the
phenomenological constraints and the consideration of dark matter
lead the theory toward one with the gauge-Yukawa unification.}

Since the gauge-Yukawa unification scenarios require the correlation
among SUSY mass parameters, precise measurement of the SUSY parameters
will be important and necessary in order to probe the gauge-Yukawa
unifications.

\section*{Acknowledgements}
We thank W.A. Bardeen, R.~Derm\' \i \v sek, B. Dobrescu, T. Li, J. Lykken and
Y. Nomura  for useful discussions. S.N.  thanks the Fermilab Theory Group
for warm hospitality and support during the completion of this work.  K.T. was
supported in part by the U.S. Department of Energy, and Y.M. was
by the Natual Sciences and Engineering Research Council of Canada.
The work of I.G. and S.N. are supported in part by DOE Grants
DE-FG03-98ER-41076 and  DE-FG02-01ER-45684.


\end{document}